\documentclass[aps,prd,twocolumn,twoside,superscriptaddress,floatfix]{revtex4-1}
\usepackage{amsmath}
\usepackage{bm}

\newcommand\nhat{\hat{\mathbf n}}

\newcommand\beq{\begin{equation}}
\newcommand\eeq{\end{equation}}
\newcommand\beqn{\begin{eqnarray}}
\newcommand\eeqn{\end{eqnarray}}

\newcommand{\ba}{\begin{eqnarray}}
\newcommand{\ea}{\end{eqnarray}}
\newcommand{\be}{\begin{equation}}
\newcommand{\ee}{\end{equation}}

\newcommand\lsim{\mathrel{\rlap{\lower4pt\hbox{\hskip1pt$\sim$}}
        \raise1pt\hbox{$<$}}}
\newcommand\gsim{\mathrel{\rlap{\lower4pt\hbox{\hskip1pt$\sim$}}
        \raise1pt\hbox{$>$}}}

\usepackage{graphicx}
\usepackage[large]{subfigure}
\usepackage{amssymb, amsmath}
\usepackage[amssymb]{SIunits}
\usepackage{epstopdf}
\usepackage{aas_macros}
\usepackage{natbib}

\begin{document}

\title{The Atacama Cosmology Telescope: Cross-Correlation of CMB Lensing and Quasars}
\author{Blake D.~Sherwin}
\email{bsherwin@princeton.edu}
\affiliation{Dept.~of Physics,
Princeton University, Princeton, NJ, USA 08544}
 \author{Sudeep~Das}
 \affiliation{BCCP, Dept.~of Physics, University of California, Berkeley, CA, USA 94720}\affiliation{Dept.~of Physics,
Princeton University, Princeton, NJ, USA 08544}\affiliation{Dept.~of Astrophysical Sciences, Peyton Hall, Princeton University, Princeton, NJ USA 08544}
\author{Amir~Hajian}\affiliation{CITA, University of
Toronto, Toronto, ON, Canada M5S 3H8}
 \author{Graeme~Addison}
 \affiliation{Dept.~of Astrophysics, Oxford University, Oxford, 
UK OX1 3RH}
 \author{J.~Richard~Bond}
 \affiliation{CITA, University of
Toronto, Toronto, ON, Canada M5S 3H8}
\author{Devin Crichton}
\affiliation{Dept.~of Physics and Astronomy, The Johns Hopkins University, Baltimore, MD 21218-2686}
\author{Mark~J.~Devlin}\affiliation{Dept.~of Physics and Astronomy, University of
Pennsylvania, Philadelphia, PA, USA 19104}
\author{Joanna~Dunkley}\affiliation{Dept.~of Astrophysics, Oxford University, Oxford, 
UK OX1 3RH}
\author{Megan B.~Gralla}
\affiliation{Dept.~of Physics and Astronomy, The Johns Hopkins University, Baltimore, MD 21218-2686}
\author{Mark~Halpern}\affiliation{Dept.~of Physics and Astronomy, University of
British Columbia, Vancouver, BC, Canada V6T 1Z4}
 \author{J.~Colin Hill}
 \affiliation{Dept.~of Astrophysical Sciences, Peyton Hall, Princeton University, Princeton, NJ USA 08544}
\author{Adam~D.~Hincks}\affiliation{CITA, University of
Toronto, Toronto, ON, Canada M5S 3H8}
\author{John~P.~Hughes}\affiliation{Dept.~of Physics and Astronomy, Rutgers, 
The State University of New Jersey, Piscataway, NJ USA 08854-8019}
\author{Kevin~Huffenberger}
\affiliation{Department of Physics, University of Miami, Coral Gables, Florida 33146}
\author{Ren\'ee~Hlozek}\affiliation{Dept.~of Astrophysical Sciences, Peyton Hall, Princeton University, Princeton, NJ USA 08544}
\author{Arthur~Kosowsky}\affiliation{Dept.~of Physics and Astronomy, University of Pittsburgh, 
Pittsburgh, PA, USA 15260}
\author{Thibaut~Louis}
 \affiliation{Dept.~of Astrophysics, Oxford University, Oxford, 
UK OX1 3RH}
\author{Tobias~A.~Marriage}\affiliation{Dept.~of Physics and Astronomy, The Johns Hopkins University, Baltimore, MD 21218-2686}\affiliation{Dept.~of Astrophysical Sciences, Peyton Hall, 
Princeton University, Princeton, NJ USA 08544}
\author{Danica~Marsden}\affiliation{Department of Physics, University of California, Santa Barbara, CA 93106 US}
\author{Felipe~Menanteau}\affiliation{Dept.~of Physics and Astronomy, Rutgers, 
The State University of New Jersey, Piscataway, NJ USA 08854-8019}
\author{Kavilan~Moodley}\affiliation{Astrophysics and Cosmology Research Unit, Univ. of KwaZulu-Natal, Durban, 4041,
South Africa}
\author{Michael~D.~Niemack}\affiliation{NIST Quantum Devices Group, 325
Broadway Mailcode 817.03, Boulder, CO, USA 80305}\affiliation{Dept.~of Physics,
Princeton University, Princeton, NJ, USA 08544}
\author{Lyman~A.~Page}\affiliation{Dept.~of Physics,
Princeton University, Princeton, NJ, USA 08544}
\author{Erik~D.~Reese}\affiliation{Dept.~of Physics and Astronomy, University of
Pennsylvania, Philadelphia, PA, USA 19104}
\author{Neelima~Sehgal}\affiliation{Dept.~of Astrophysical Sciences, Peyton Hall, Princeton University, Princeton, NJ USA 08544}
\author{Jon~Sievers}\affiliation{Dept.~of Physics,
Princeton University, Princeton, NJ, USA 08544}
\author{Crist\'obal Sif\'on}
\affiliation{Departamento de Astronom\'ia y Astrof\'isica, Facultad de F\'isica, Pontificia Universidad Cat\'olica de Chile, Casilla 306, Santiago 22, Chile}
\author{David~N.~Spergel}\affiliation{Dept.~of Astrophysical Sciences, Peyton Hall, 
Princeton University, Princeton, NJ USA 08544}
\author{Suzanne~T.~Staggs}\affiliation{Dept.~of Physics,
Princeton University, Princeton, NJ, USA 08544}
\author{Eric~R.~Switzer}\affiliation{Kavli Institute for Cosmological Physics, 
5620 South Ellis Ave., Chicago, IL, USA 60637}\affiliation{Dept.~of Physics,
Princeton University, Princeton, NJ, USA 08544}
\author{Ed~Wollack}\affiliation{Code 553/665, NASA/Goddard Space Flight Center,
Greenbelt, MD, USA 20771}


\begin{abstract}
We measure the cross-correlation of Atacama Cosmology Telescope CMB lensing convergence maps with quasar maps made from the Sloan Digital Sky Survey DR8 SDSS-XDQSO photometric catalog. The CMB lensing-quasar cross-power spectrum is detected for the first time at a significance of $3.8 \sigma$, which directly confirms that the quasar distribution traces the mass distribution at high redshifts $z>1$. Our detection passes a number of null tests and systematic checks. Using this cross-power spectrum, we measure the amplitude of the linear quasar bias assuming a template for its redshift dependence, and find the amplitude to be consistent with an earlier measurement from clustering; at redshift $z \approx 1.4$, the peak of the distribution of quasars in our maps, our measurement corresponds to a bias of $b = 2.5\pm 0.6$. With the signal-to-noise ratio on CMB lensing measurements likely to improve by an order of magnitude over the next  few years, our results demonstrate the potential of CMB lensing cross-correlations to probe astrophysics at high redshifts.
\end{abstract}
\maketitle

\section{Introduction}

As the photons of the cosmic microwave background (CMB) travel through the universe, they are gravitationally deflected by the web of matter through which they pass. In the CMB sky we observe today, these deflections are imprinted as arcminute-scale distortions of small scale temperature fluctuations \cite{lew06,han11}. The microwave background thus contains not only information about the primordial universe at redshift $\approx 1100$, but also about the matter density fluctuations in the lower redshift universe.

The lensing deflection field $\mathbf{d}(\mathbf{\hat{n}})$ points from the direction $\mathbf{\hat{n}}$ in which a CMB photon was received to the direction from which it was emitted. This deflection field can be determined from the measured CMB because lensing changes the statistics of small scale unlensed CMB fluctuations in a characteristic way, introducing correlations between different Fourier modes. By measuring correlations between pairs of Fourier modes that would be uncorrelated in the absence of lensing, one can estimate $\mathbf{d}$ \cite{hu02} and hence the lensing convergence $\mathbf{\kappa}\equiv -\nabla \cdot \mathbf{d}/2$ (a useful quantity because it is a direct measure of the projected matter density, see Eq.~1). Using a quadratic estimator, lensing was first measured in cross-correlation with radio sources and galaxies by \cite{smi07,hir08} (using WMAP data) and in auto-correlation by \cite{das11} (using Atacama Cosmology Telescope data). More accurate measurements of both the lensing power spectrum and the lensing-galaxy cross-correlation were recently reported by the South Pole Telescope \cite{van12, ble12}.

CMB lensing measurements are a powerful cosmological probe \cite{dep09} because they are sensitive to both the growth of density fluctuations and the geometry and size of the universe, yet are relatively insensitive to both instrumental and astrophysical systematic errors \cite{das11,van12}. Lensing measurements can constrain the properties of dark energy \cite{she11}, the amplitude of density fluctuations and the masses of neutrinos \cite{les06}. They can also constrain the properties of biased tracers of the matter distribution. The focus of this paper is the cross-correlation of CMB lensing maps with one such tracer -- quasars.

Quasars are among the most luminous objects in the universe. Their immense luminosity is believed to be powered by accreting supermassive black holes \cite{sal,lyn} which reside at the center of almost every massive galaxy \cite{kor95}.  As the activity of quasars and the rate of star formation appear to be linked \cite{nan07}, they are a crucial element in our present understanding of galaxy evolution. Measurements of the relation between dark matter and the distribution of quasars can inform us about quasar properties such as the masses of the dark matter halos that host the quasars, the scatter in the quasar-halo mass relation and the quasar duty cycle (see e.g.~\cite{white}). Such measurements of quasar properties will, in turn, improve our understanding of structure formation and galaxy evolution.

Both the number density of quasars and the strength of CMB lensing in a certain direction depend on the projected dark matter density in this direction, and quasars are most common at the redshifts that produce the largest lensing deflections. This implies that the CMB lensing and quasar fields should be strongly correlated \cite{pei00}. Measuring the cross-power spectrum and comparing it to theoretical calculations, we can determine the proportionality factor which relates a fluctuation in matter density to a fluctuation in quasar number density. This proportionality factor is known as the quasar linear bias $b$ (defined as $b \equiv \delta_q / \delta$ where $\delta_q $ and $  \delta$ are the fractional spatial overdensities of quasars and matter respectively).

In this work we present the first measurement of the CMB lensing-quasar cross-power spectrum, and use it to derive a constraint on the quasar bias. The paper is organized as follows. Section II presents the theoretical background underlying the CMB lensing-quasar
cross-correlation. Section III explains how the lensing and quasar maps used in our
analysis are constructed, and describes the resulting cross-spectrum measurement.
The constraint on quasar linear bias we obtain from the cross-power spectrum is presented in section IV. 
Null tests and systematic checks are discussed in section V. All calculations assume a $\Lambda$CDM cosmology with WMAP5 parameters \cite{Komatsuetal2009} and $\sigma_8=0.819$.

\section{Theoretical Background}

Cosmological weak lensing effects can be described using the convergence field $\kappa$, which is equal to a weighted projection of the matter overdensity $\delta$ \cite{lew06} 
\beq
\kappa(\nhat)=\int_0^{z_{LS}}  dz W^\kappa (z) \delta(\eta(z) \nhat,z).
\eeq 
The relevant kernel (assuming a flat universe) is
\beq
W^\kappa(z) = \frac{3}{2 H(z)} \Omega_0 H_{0}^{2} (1+z) \eta(z) \frac{\left(\eta^{LS}-\eta(z)\right)}{\eta^{LS}}
\eeq
where $\eta(z)$ is the comoving distance to redshift $z$, $\hat{\mathbf{n}}$ is a direction on the sky, $\eta^{LS}$ is the comoving distance to the last scattering surface, $z_{LS}$ is the redshift of the last scattering surface, $H(z)$ is the Hubble parameter, and $\Omega_0$  and $H_{0}$ represent the present values of the matter density parameter and the Hubble parameter, respectively.

The fractional overdensity of quasars in a direction $\nhat$ is given by $q(\nhat)$, which -- assuming a linear bias relation between the distribution of matter and quasars -- is given by
\beq
q(\nhat)=\int_0^{z_{LS}} dz W^q(z) \delta(\eta(z) \nhat,z),
\eeq
where the kernel is
\beq
W^q(z) = \frac{b(z)\frac{dN}{dz}}{ \left[\int dz'\frac{dN}{dz'}\right]} +  \frac{3}{2 H(z)} \Omega_0 H_{0}^{2} (1+z)g(z)(5s-2)
\eeq
and
\beq
g(z) = \eta(z) \int_z^{z_{LS}} dz' (1-\eta(z)/\eta(z')) \frac{\frac{dN}{dz'}}{ \left[\int dz''\frac{dN}{dz''}\right]} 
\eeq
(see \cite{pei00}).  Here $b$ is the linear bias, $dN/dz$ is the redshift distribution of quasars, and $s$ is the variation of the number counts of quasars $N(<m)$ with the limiting magnitude $m$ at the faint limit of the quasar catalog, $s\equiv d \log_{10} N / d m$. The second term in Eq.~4 is the magnification bias: the change in source density in a certain direction induced by lensing magnification. For the quasar catalog used in this work (which has $s\approx0.043$) this term is significantly smaller than the first term ($\approx 15\%$ of its magnitude) and is negative.

Combining Eqs.~1 and 3 gives the expected lensing-quasar cross-power spectrum in the Limber approximation \cite{lim54}:
\beq
C_\ell^{\kappa q} = \int \frac{dz H(z)}{ \eta^2(z)} W^\kappa(z) W^q(z) P(k = \ell/\eta(z),z),
\eeq
where $P(k,z)$ is the matter power spectrum at wavenumber $k$ and redshift $z$.

\section{Cross-correlating CMB Lensing and Quasars}
\subsection{The ACT CMB Lensing Convergence Maps}
The lensing convergence fields used in our analysis are reconstructed from CMB temperature maps made by the Atacama Cosmology Telescope (ACT) \cite{fow07,dun12,swe11}, a 6m telescope operating in the Atacama desert in Chile. These CMB temperature maps are obtained from observations made during 2008-2010 in the 148 GHz frequency band and calibrated as in \cite{haj11}. The maps consist of six patches, each of size $3\times 18$ degrees, in the Sloan Digital Sky Survey (SDSS) Stripe 82 region \cite{ann11}, with a map-averaged white noise level of 21 $\mu$K arcmin. Radio and IR point sources as well as Sunyaev-Zel'dovich (SZ) clusters detected with a matched filter at a signal-to-noise ratio greater than 5 as in \cite{das11} are masked and inpainted using the methods of \cite{lou11}.

The convergence maps are reconstructed from the CMB temperature maps as in \cite{das11} using the minimum variance quadratic estimator procedure described in \cite{hu02}. The estimator works as follows. While the unlensed CMB is statistically isotropic, any lensing deflection imprints a preferred direction into the statistical properties of the CMB. This corresponds mathematically to the fact that formerly uncorrelated modes of the isotropic unlensed CMB temperature field are correlated by lensing, with the correlation proportional to the lensing deflection. We can hence estimate the lensing convergence by measuring the correlation between pairs of Fourier modes using a quadratic estimator:
\beq
\hat{\kappa}(\mathbf{L}) = N(\mathbf{L}) \int \mathrm{d}^2 \mathbf{l}~f(\mathbf{L},\mathbf{l})~T(\mathbf{l}) T(\mathbf{L}-\mathbf{l}) .
\eeq
Here $\mathbf{l},\mathbf{L}$ are Fourier space coordinates, $N$ is the normalization function (which ensures that the estimator is unbiased) and $f$ is a weighting such that the
signal-to-noise ratio on the reconstructed convergence is maximized in
the case of isotropic noise with full sky coverage (see \cite{hu02} for the details of these functions). The weighting uses a smoothed version of the observed anisotropic noise power. To calculate the normalization function we multiply a first-approximation normalization which uses the data power spectrum by a small $L$-dependent correction factor. This factor is obtained from Monte-Carlo simulations by requiring that on average the cross-power of the reconstructed convergence with the true simulation convergence be equal to the true convergence power spectrum. The simulated maps used in this Monte-Carlo calculation are constructed to match ACT data in both signal and noise properties as in \cite{das11}. In our lensing reconstruction, we use a wider range of scales in the temperature map than in \cite{das11}, filtering out modes below $\ell =500$ and above $\ell=4000$. As can be seen in Eq.~7, lensing information at a scale $\ell$ is obtained from two temperature modes \emph{separated} by $\ell$, so that this filtering does not preclude us from obtaining small-$\ell$ lensing modes. Intuitively, this is because we deduce the distribution of large scale lenses from the distortion of small scale temperature fluctuations.

We subtract from the reconstructed ACT convergence maps a simulated mean field map $\langle \hat{\kappa} \rangle$, obtained from 480 realizations of simulated reconstructed lensing maps. This map is non-zero due to correlations induced by window functions and noise which, to the convergence estimator, appears as a small spurious lensing signal which must be subtracted.

We can thus estimate the CMB lensing-quasar cross power spectrum by calculating
\beq
C_\ell^{\kappa q} = \left \langle \mathrm{Re} [( \hat{\kappa}(\mathbf l)-\langle \hat{\kappa} \rangle(\mathbf l))^*q(\mathbf l)] \right \rangle_{\mathbf{l} \in \ell}
\eeq
where the outer average is over all pixels with Fourier coordinates $\mathbf{l}$ which lie within the bandpower corresponding to $\ell$.

\begin{figure}[!h]
\label{fig.qsos}
    \includegraphics[width=4.1in]{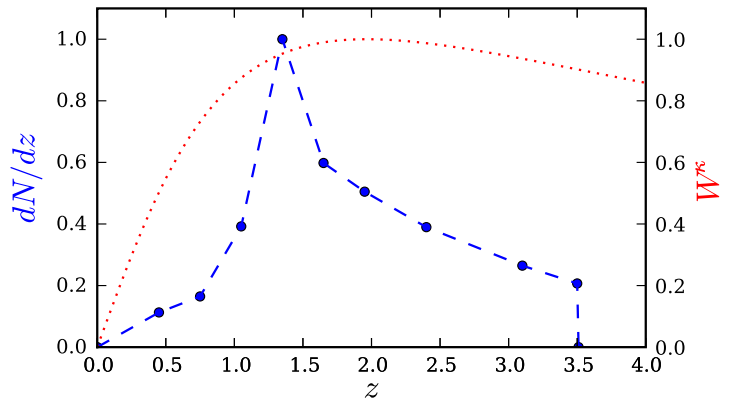}
    \caption{The redshift distribution of SDSS quasars used to construct our maps of fractional quasar overdensity, normalized to a unit maximum. The corresponding redshift bins are shown with blue filled circles; they are interpolated to give the continuous curve used in our theory calculations (blue dashed line). For comparison, the red dotted line shows the lensing kernel $W^{\kappa}(z)$, again normalized to a unit maximum.}
\end{figure}

\subsection{The SDSS Quasar Maps}
In this work we use the SDSS-XDQSO photometric quasar catalog \cite{bov11,bov12}, extracted from the SDSS Data Release 8. The analysis used in this catalog separates the population of quasars and foreground stars using a probabilistic model in flux space. This analysis assigns a probability of being a quasar to every point-source with de-reddened i-band magnitude between 17.75
and 22.45 mag in the SDSS imaging. Though the catalog extends out to $z>4$, we do not use the highest-redshift sources with redshifts $z>3.5$. This reduces the shot noise error on our measurement, as there are very few such sources. Using this catalog, we construct a map of the fractional overdensity $q$ of quasars across the same $324$ square degrees on which we perform our lensing reconstruction. We include in our quasar maps, with unit weight, all point sources with a greater than $p=0.68$ probability of being a quasar. As the probability distribution is non-uniform, the residual level of stellar contamination can be calculated from the catalog probabilities to be $6\%$ (we discuss later how this is accounted for in our theoretical calculations). A spectroscopic quasar sample at high redshifts ($2.2<z<3.5$) does show a contamination fraction which is $\approx 15\%$ larger than that predicted by the catalog (see \cite{bov12}); however, the majority of the quasars we consider are at lower redshift where the XDQSO model estimates should be significantly more accurate (high redshift quasar selection is less accurate because at $z \approx 2.8$ the quasar and stellar loci cross in color space). We neglect the error on the stellar contamination fraction, which is any case smaller than the statistical error (and would only reduce the measured signal, not increase it). The area covered by our quasar maps contains on average 75 quasars per square degree. The redshift distribution of the quasars in our maps is shown in Fig.~1, along with the CMB lensing kernel, which has a very similar redshift distribution.

\subsection{The CMB Lensing - Quasar Cross-Power Spectrum}
The cross-power spectrum of the ACT CMB lensing
maps and the SDSS quasar maps is shown in Fig.~2.

\begin{figure}[!h]
\label{fig.filter}
    \includegraphics[width=4.1in]{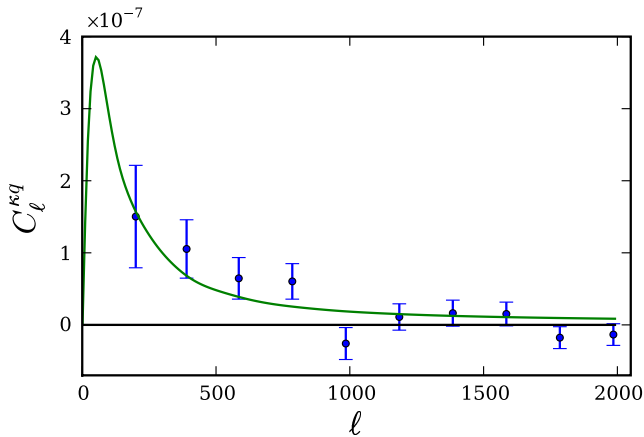}
    \caption{The CMB lensing - quasar density cross-power spectrum, with the data points shown in blue (the covariance between different data points is negligible). The significance of the detection of the cross-spectrum is $3.8 \sigma$. The green solid line is a theory line calculated assuming the fiducial bias amplitude. This theory line is reduced by $6\%$ to account for the expected level of stellar contamination of the quasar sample.}
\end{figure}
\begin{figure}[h]
\label{fig.filter}
    \includegraphics[width=4.1in]{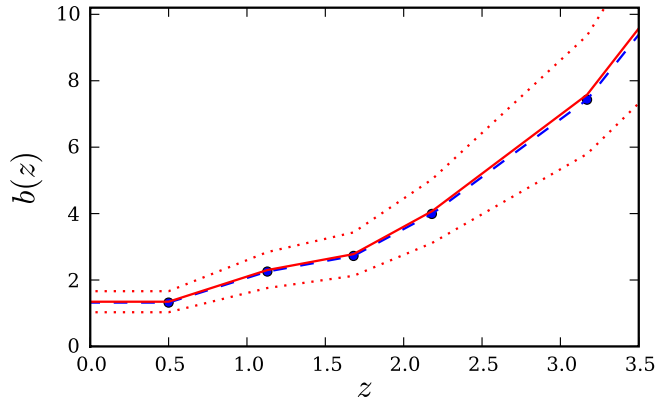}
    \caption{Blue dashed line: fiducial quasar bias template (interpolated from the data points of \cite{she09}), used in the theoretical calculation of the CMB lensing - quasar cross-power spectrum. Red solid line: the bias amplitude ($b/b_{\mathrm{fid}}=1.02$) best fit by the measured cross-power spectrum; red dashed lines: the $\pm 1 \sigma$ error ranges of this amplitude. Note that only one degree of freedom is constrained: the overall bias amplitude of an assumed redshift evolution.}
\end{figure}

The error bars on the data points are calculated theoretically as proportional to $\sqrt{C_\ell^{\kappa \kappa} C_\ell^{q q}}$, with an additional factor calculated from the number of independent pixels in the bin corresponding to each data point. The spectra used in this calculation are the full data spectra which include both signal and noise (including Poisson noise). Bootstrap error estimates from splits of our data are consistent with this calculation. For comparison, we also calculate error bars using Monte-Carlo methods, cross correlating 480 realizations of simulated reconstructed lensing maps (containing signal and realistic noise) with the quasar data maps. Both methods assume the two maps are uncorrelated; such calculations are very good approximations to the true error on the cross-correlation because both maps are noisy so that $C_\ell^{\kappa \kappa} C_\ell^{q q} \gg (C_\ell^{\kappa q})^2$. The error bars obtained from Monte-Carlo methods are nearly indistinguishable from the theory error bars, and lead to the same detection significance. (We also verified that replacing the quasar data maps with 480 simulated maps with the same number of randomly distributed sources leads to similar, though slightly smaller, errors.) The Monte-Carlo estimates of the errors also allow us to calculate the full covariance matrix. The off-diagonal elements are negligible compared to the diagonal elements; for every bin, the covariance between neighboring bins was less than $4\%$ of the bin autocorrelation. We thus neglect covariance between different data points in our analysis.

Also shown in Fig.~2 is a theoretical calculation of the expected cross-power spectrum obtained from Eq.~6. In this calculation the matter power spectrum was computed using the CAMB software \cite{camb}. The non-linear (HALOFIT, \citep{halofit}) matter power spectrum was used; however, using a linear matter power spectrum instead only slightly changed the computed cross-spectrum (as most of the signal arises from angular scales corresponding to linear scales in the matter power spectrum, where the linear and non-linear matter power spectra hardly differ). We use the quasar redshift distribution as shown in Fig.~1 in this calculation. As the integration kernel is slowly varying, the theory curve is insensitive to the binning and interpolation of this redshift distribution. A fiducial bias model for this calculation is obtained by interpolating the central measured bias values of \cite{she09} (averaging the values obtained with and without the inclusion of negative points in the correlation function). The fractional error on these central bias values is below $10\%$ at low redshifts $z<2$, but rises to $\sim 20\%$ at $z\sim 4-5$.  The values were obtained from measurements of the amplitude of the quasar correlation function (which is sensitive to the bias) for an SDSS spectroscopic quasar sample. This fiducial bias model is shown in Fig.~3. As the theoretical cross power spectrum does not depend strongly on the detailed form of the bias model, we use this measurement as a convenient fiducial template, though the spectroscopic catalog used in this measurement does not extend to as faint a magnitude as the photometric catalog we use to make quasar maps. Despite this, the quasar power spectrum predicted by this fiducial bias model is consistent with the power spectrum of our quasar maps. The calculated theoretical cross-power spectrum is reduced by $6\%$ to account for stellar contamination; while stars are uncorrelated with lensing, they contribute to the average density of sources, and so cause us to calculate a fractional quasar overdensity that is $6\%$ too small.

The data fit the theory curve (which assumes the fiducial bias model) well, with a chi-squared value for this curve of $\chi^2_{\mathrm{theory}}  = 13.2$ for 10 degrees of freedom. We obtain the significance of our detection of the cross-power spectrum by calculating $\sqrt{\chi^2_{\mathrm{null}} -\chi^2_{\mathrm{theory}} }$, where $\chi^2_{\mathrm{null}} $ is calculated for the null line. The significance of the detection is found to be $3.8 \sigma$.

\section{A Constraint on the Quasar Bias}
We calculate a constraint on the linear bias of quasars from the lensing-quasar cross-power spectrum. To do so we assume a bias template of the fiducial shape shown in Fig.~3, but rescaled by a constant factor for all redshifts. We calculate the likelihood as a function of this scaling parameter $b/b_{\mathrm{fid}}$ and plot it in Fig.~4. Our result, $b/b_{\mathrm{fid}}=1.02\pm0.24$, is consistent with the fiducial bias model, i.e.~a value of unity. (Due to the small negative magnification bias, $b/b_{\mathrm{fid}}=0$ does not correspond exactly to the null line.) As the redshift distribution is peaked at $z=1.4$, $b/b_{\mathrm{fid}}$ can be interpreted as approximately parametrizing the amplitude of the bias at redshift $1.4$. From this interpretation we obtain a value of the bias at $z\approx1.4$ of $b = 2.5\pm 0.6$, which is in good agreement with previous measurements from quasar clustering \cite{she09}. We can associate this bias with a host halo mass $M_{200}$: using the bias model of \cite{tin10}, we obtain a halo mass of $\log_{10}(M_{200}/M_\odot) = 12.9^{+0.3}_{-0.5}$, consistent with previous estimates \cite{she09, white}. We also verify that the cross-power calculated using only a low ($z < 2.2$) or high ($z>2.2$) redshift quasar sub-sample is consistent with the bias given by the fiducial model; however, we defer a detailed calculation of bias constraints using multiple quasar sub-samples to future work with a higher signal-to-noise ratio.

\begin{figure}[!h]
\label{fig.filter}
    \includegraphics[width=4.1in]{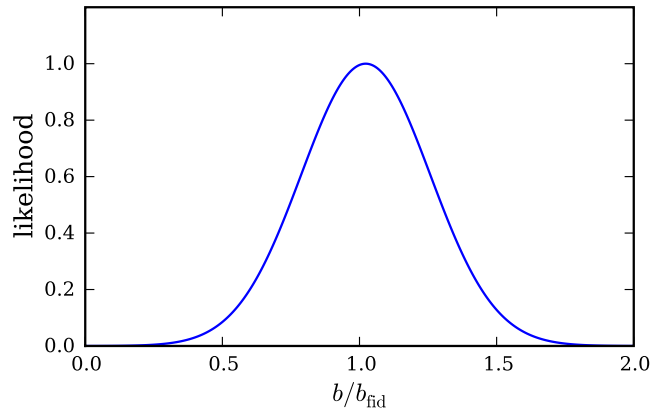}
    \caption{Likelihood as a function of quasar bias divided by the fiducial bias, $b/b_{\mathrm{fid}}$ (we assume that the shape of the redshift dependence is constant and has the fiducial form of Fig.~3, and modify the amplitude of the bias function to calculate this likelihood.) Interpreting our measurement of $b/b_{\mathrm{fid}}=1.02\pm0.24$ as a bias at $z\approx1.4$ (the peak in the quasar distribution), we obtain $b=2.5\pm0.6$ at this redshift.}
\end{figure}

\section{Testing the power spectrum}

\subsection{Null Tests}
We check our result and our pipeline with a number of null tests. In a simple first test, we cross-correlate the quasar distribution in one part of the sky with the lensing convergence in another; as seen in Fig.~5, the results are consistent with null as expected, with $\chi^2 = 6.5$ for 10 degrees of freedom for a fit to null. A more sophisticated test is to calculate the cross-correlation of the quasar maps with the curl component of the lensing deflection (this differs from the convergence reconstructed earlier, which is gradient-like). The reconstructed curl map is expected to be zero (though it should contain reconstruction noise), and hence the cross-correlation with the quasar maps should be zero as well. We reconstruct the curl component of the estimator as in \cite{van12} (though keeping the normalization and filters unchanged from the earlier convergence reconstruction, and simply replacing the dot product in $f(\mathbf{L},\mathbf{l})$ of Eq.~7 with a cross product projected onto the $\bf{\hat{ l}}_x \times \bf{\hat{ l}}_y$ direction). The cross-correlation of the lensing curl component with the quasar maps is also shown in Fig.~5. The error bars are calculated from theory as before. As expected, this test is consistent with a null result, with a (somewhat low) value of $\chi^2 = 3.5$ for 10 degrees of freedom. 

\begin{figure}[!h]
\label{fig.filter}
    \includegraphics[width=4.1in]{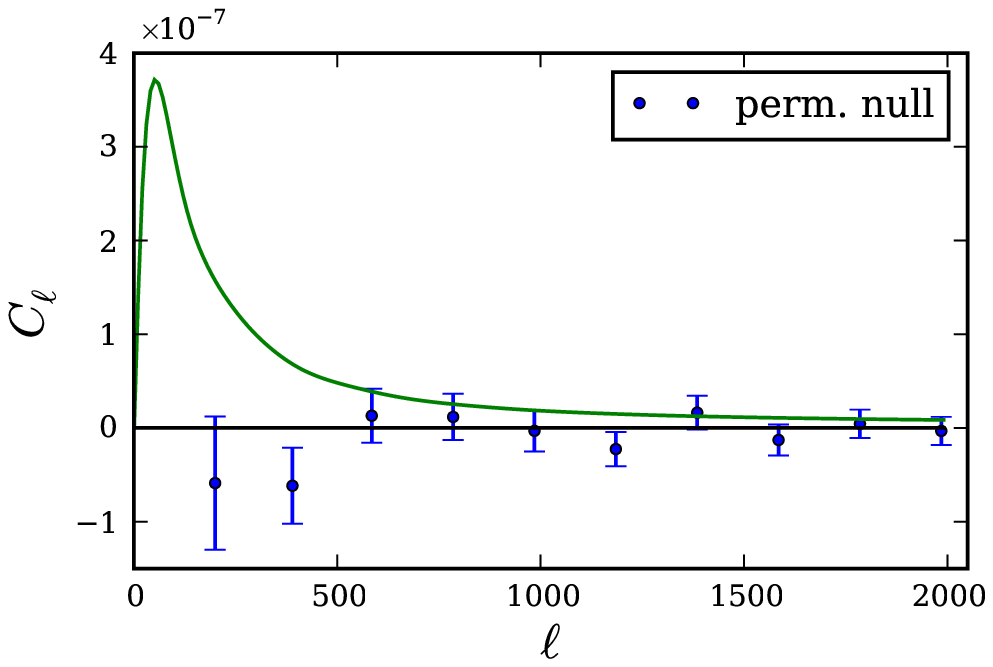}
    \includegraphics[width=4.1in]{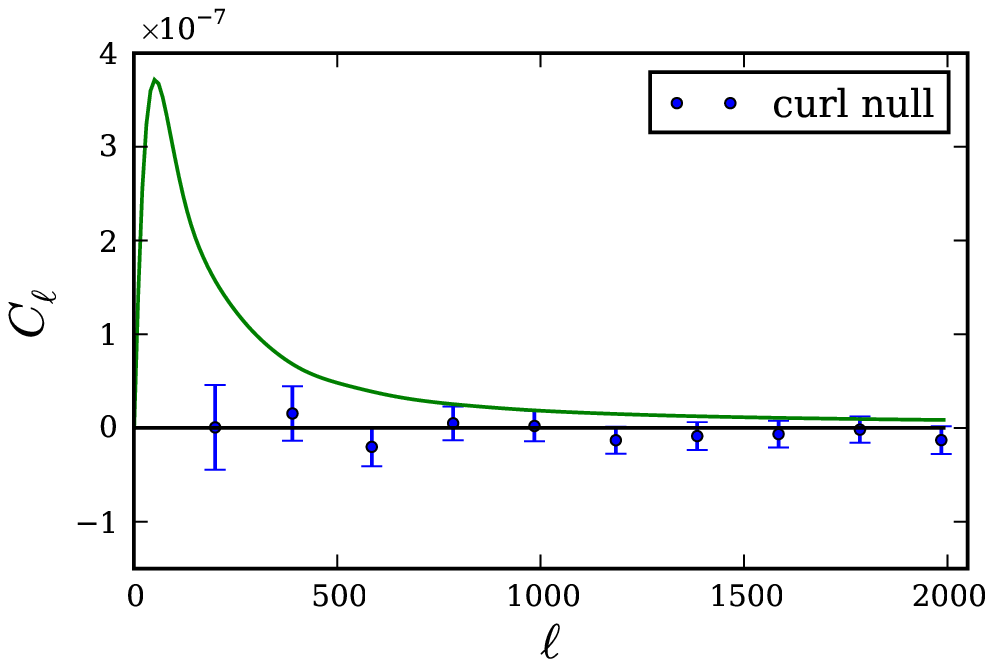}
    \caption{Two successful null tests, both consistent with zero. Upper panel: the cross-power spectrum of quasar and lensing maps covering different parts of the sky (permutation null test). Lower panel: the cross-power spectrum of the reconstructed curl component of the lensing signal with the quasar maps (curl null test).}
\end{figure}

\subsection{Estimating Potential Systematic Contamination}

We estimate the magnitude of what are expected to be the largest contaminants: infrared (IR) sources, SZ clusters and galactic cirrus, which contribute flux to the CMB temperature maps. (The level of radio source power is much smaller as we can resolve and mask such sources down to low flux levels.) Contamination of the cross-power spectrum is conceivable because the sources of IR and SZ signal trace the underlying matter field as the quasars do, and galactic cirrus could reduce observed quasar counts by extinction. As explained in \cite{ble12}, any IR or SZ contamination would appear as a negative bias, as a large IR or SZ signal in a certain direction increases local gradients; the lensing estimator falsely interprets this as a signature of demagnification of the CMB and hence estimates a spurious underdensity in this direction.

To obtain an estimate of the contribution of IR and SZ contamination to the measured cross-power spectrum, we construct simulated quasar maps which are correlated with the IR and SZ maps of the simulations of \cite{seh10}. The quasar maps are constructed by randomly populating all halo positions (listed for all halos with masses greater than $6.8\times 10^{12} M_{\odot}$ in the catalog supplied with \cite{seh10}) with quasars. In populating the halos we use a redshift-dependent probability of occupation such that the final simulated quasar map has the correct redshift distribution. Cross-correlating this quasar map with the true convergence maps of the same simulation, the signal is consistent with the theory line of Fig.~2, which confirms that our simple simulation has approximately the correct bias (this is due to the average mass in the halo catalog being similar to the typical halo mass of a linear bias model consistent with the fiducial model).

The level of systematic contamination in our estimator can be obtained
from these simulated quasar, IR, and SZ maps. Keeping the same
filtering and normalization as in our data, we reconstruct the IR
contaminant to the lensing signal by replacing the temperature maps in
Eq.~7 with the simulated IR maps (which we rescaled using an appropriate factor as in \cite{van12} to match more recent constraints on IR source flux). The cross-spectrum of the
resulting map with the simulated quasar maps gives a negative spurious signal which is $\approx 7\%$ of
the theoretical prediction for the lensing-quasar cross-correlation.
Repeating this analysis with the thermal SZ simulations gives a similar negative
contamination of order $\approx 5\%$ of theory. (The analysis should overestimate the contamination, as the simulated quasars are placed exactly in the centers of the same halos that source the SZ and IR signal, neglecting any mis-centering effects.)
Any systematic error in our measurement of the lensing-quasar cross-power spectrum
due to contamination from both IR and SZ sources should thus be
significantly smaller than the size of the statistical error. In addition, the fact that these contamination signals are negative means that our detection of a positive lensing-quasar cross-power spectrum cannot be due to such systematics.

Finally, we estimate the level of contamination from galactic cirrus using the dust maps of \cite{dfs}. We subtract a map of the signal at 148GHz induced by the dust (obtained from \cite{seh10}) from the ACT temperature data, reconstruct lensing, and re-estimate the lensing-quasar cross-power spectrum. We find that the change in the cross-power spectrum is very small, of order 3\% of the theoretical prediction for the lensing-quasar cross-spectrum. (This is unsurprising, as in our analysis large scale power below $\ell=500$ has been filtered out of the temperature maps.) Contamination of the cross-power by galactic cirrus is thus negligible.

\section{Summary and Conclusions}
In this work we measure the cross-correlation between ACT CMB lensing maps and maps of the quasar distribution made from the SDSS-XDQSO catalog. We detect the cross-power spectrum at $3.8 \sigma$ significance, directly confirming that quasars trace mass. We check our detection with null tests including a cross-correlation of quasars with the reconstructed curl component of lensing, which is found to be null as expected. Potential systematic contamination is estimated and found to be negligible. From our detection we estimate the quasar bias.  We measure $b/b_{\mathrm{fid}}=1.02\pm0.24$; interpreting this as a bias at $z\approx1.4$ (the peak in the quasar distribution), we obtain $b=2.5\pm0.6$ at this redshift (which corresponds to a host halo mass of $\log_{10}(M_{200}/M_\odot) = 12.9^{+0.3}_{-0.5}$). Unlike measurements from clustering, this lensing measurement involves a direct comparison of the quasar distribution with the mass distribution, with little modeling required.

The study of high-redshift mass tracers with CMB lensing is a new field. In the next few years, the signal-to-noise ratio on lensing measurements should improve by an order of magnitude with data from experiments such as Planck, ACTPol and SPTPol \cite{planck,nie10,mcm09}. ACTPol in particular should provide high signal-to-noise ratio lensing measurements which have considerable overlap with SDSS quasar fields. Higher signal-to-noise will allow constraints on quasar biases as a function of redshift, luminosity, color or other properties and will thus provide a wealth of information on the properties of quasars and the halos that host them. More precise bias measurements of both quasars and galaxies will also allow tests of dark energy properties \cite{das09} and modified gravity \cite{acq08}. This work lies at the beginning of an exciting research program: the study of astrophysics and cosmology with CMB lensing cross-correlations.

\begin{acknowledgments}
We thank Jo Bovy and Michael Strauss for discussions and helpful comments on the draft, and acknowledge useful discussions with Kendrick Smith and Alex van Engelen. This work was supported by the U.S.\ NSF through awards AST-0408698, PHY-0355328, AST-0707731 and PIRE-0507768, as well as by Princeton Univ., the Univ. of Pennsylvania,
FONDAP, Basal, Centre AIUC, RCUK Fellowship (JD),  NASA grant NNX08AH30G (SD, AH, TM), NSERC PGSD (ADH), NSF PFC grant PHY-0114422 (ES),  KICP Fellowship (ES), SLAC no. DE-AC3-76SF00515 (NS), ERC grant 259505 (JD), BCCP (SD), and the NSF GRFP (BDS, BLS). We thank  B.\ Berger, R.\ Escribano, T.\ Evans, D.\ Faber, P.\ Gallardo, A.\ Gomez, M.\ Gordon, D.\ Holtz, M.\ McLaren, W.\ Page, R.\ Plimpton, D.\ Sanchez, O.\ Stryzak, M.\ Uehara, and Astro-Norte for assistance with ACT. ACT operates in the Parque Astron\'{o}mico Atacama in northern Chile under the auspices of Programa de Astronom\'{i}a, a program of the Comisi\'{o}n Nacional de Investigaci\'{o}n Cient\'{i}fica y Tecnol\'{o}gica de Chile (CONICYT). Computations were performed on the GPC supercomputer at the SciNet HPC Consortium. SciNet is funded by the CFI under the auspices of Compute Canada, the Government of Ontario, the Ontario Research Fund -- Research Excellence; and the University of Toronto.
\end{acknowledgments}

\end{document}